# Thought experiments that shed light on the nature of optical linear and angular momenta

Masud Mansuripur

College of Optical Sciences, The University of Arizona, Tucson, Arizona 85721

**Abstract**. Invoking tools and techniques from elementary theories of classical electrodynamics and special relativity, we analyze some of the thought experiments that have contributed substantively to the conceptual development and understanding of the linear and angular momenta of light.

**1. Introduction**. It is well known that electromagnetic (EM) waves, much like material objects, can possess energy as well as linear and angular momenta,[1-5] and that the angular momentum can be of the spin variety (rooted in circular polarization) or of the orbital type (due to phase vorticity).[6-8] The classical theory of electrodynamics can, of course, account for the observed phenomena involving the energy and momenta of the EM field. Nevertheless, it is instructive to try to deduce some of the fundamental properties of the EM field from relatively simple thought experiments that draw on a general knowledge of classical mechanics, thermodynamics, theory of relativity, and the standard conservation laws of physics. The purpose of the present paper is to introduce the reader to some of the more interesting thought experiments — originally developed by the likes of Hendrik Lorentz and Albert Einstein — that have played important roles in the historical development of the field and contributed to our current understanding of the mechanical aspects of electromagnetic radiation.

In Sec.2, we lay the groundwork by examining the EM energy and linear momentum of a simple packet of light, typically referred to as a "light bullet." Section 3 invokes the fundamental idea behind the equivalence of mass and energy, and provides the basis for our discussion of the Einstein-box thought experiment later in Sec.6. We also use the setup of Sec.3 to argue that the EM angular momentum $\boldsymbol{L}_{\text{EM}}$ of a photon cannot depend on the oscillation frequency $\omega$ of its EM field.

In Sec.4, we look at the reflection of a plane-wave from a perfect reflector, and show how to relate the linear momentum density $\boldsymbol{p}_{\text{EM}}(\boldsymbol{r},t)$ of the EM field to its local Poynting vector $\boldsymbol{S}(\boldsymbol{r},t)$. The case of a thin, current-carrying sheet radiating circularly polarized light into its surrounding space is considered in Sec.5, where, among other things, it will be shown that the ratio of radiated energy to radiated angular momentum is proportional to the oscillation frequency of the field. This will be a reaffirmation of a conclusion reached in Sec.3 (based on relativistic arguments) that a photon's angular momentum $\boldsymbol{L}_{\text{EM}}$ cannot depend on its oscillation frequency $\omega$.

A slightly modified version of the classical Einstein-box thought experiment is the subject of Sec.6, where the magnitude $p_{\text{EM}}$ of the EM momentum of a light pulse in vacuum is shown to be equal to the pulse energy $\mathcal{E}$ divided by the speed $c$ of light in vacuum. A variation on the theme of the Einstein-box thought experiment, due to Balazs, reveals that the EM part of the linear momentum of a wavepacket inside a transparent host medium is given by the linear momentum of the same pulse in vacuum divided by the group refractive index $n_g$ of the host, that is, $p_{\text{EM}} = \mathcal{E}/(n_g c)$. Section 7 is devoted to a detailed mathematical description of various aspects of the Balazs thought experiment. The paper closes with a brief summary of its main results in Sec.8.

**2. A bullet of light**. Consider an EM wavepacket of duration $\tau$ and cross-sectional area $A$ propagating in free space along the $z$-axis, as shown in Fig.1. The electric and magnetic fields of the packet (ignoring the tapering of the fields near the boundaries) are given by

$$\boldsymbol{E}(\boldsymbol{r},t) = E_{x0}\hat{\boldsymbol{x}}\exp[\mathrm{i}(k_0 z - \omega t)]. \tag{1a}$$

$$\boldsymbol{H}(\boldsymbol{r},t) = (E_{x0}/Z_0)\hat{\boldsymbol{y}}\exp[\mathrm{i}(k_0 z - \omega t)]. \tag{1b}$$

In the above equations, $E_{xo}$ is the amplitude of the incident $E$-field, $\omega$ is the angular frequency of the monochromatic wave, $k_o = \omega/c = 2\pi/\lambda_o$ is the magnitude of the wave-vector (also known as the wave-number), $\lambda_o$ is the wavelength, $c = 1/\sqrt{\mu_o\varepsilon_o}$ is the speed of light in vacuum, and $Z_o = \sqrt{\mu_o/\varepsilon_o}$ is the impedance of free space — $\mu_o$ and $\varepsilon_o$ being the permeability and permittivity of free space.

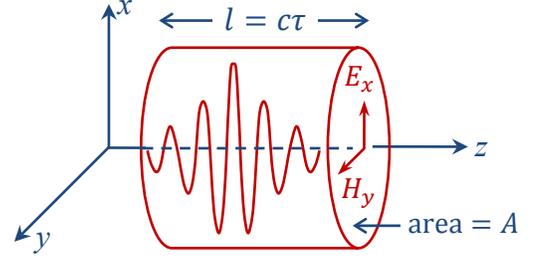

**Fig. 1**. A quasi-monochromatic light pulse of duration $\tau$ and cross-sectional area $A$ propagates along the $z$-axis. The light beam is linearly polarized, with its $E$-field being along the $x$-axis and $H$-field along the $y$-axis. The pulse has energy $\mathcal{E}$ and electromagnetic linear momentum $\boldsymbol{p}_{\text{EM}} = (\mathcal{E}/c)\hat{\boldsymbol{z}}$.

The energy content of the pulse is the sum of its $E$-field and $H$-field energy densities integrated over the volume of the pulse, that is,

$$\mathcal{E}(t) = \iiint (\tfrac{1}{2}\varepsilon_o E_{xo}^2 + \tfrac{1}{2}\mu_o H_{yo}^2)\cos^2(k_o z - \omega t)\,\mathrm{d}x\mathrm{d}y\mathrm{d}z = \tfrac{1}{2}\varepsilon_o E_{xo}^2 A c\tau. \tag{2}$$

Here, it is being assumed that the dimensions of the wavepacket are much larger than its wavelength $\lambda_o$. The same result may also be obtained by integrating the Poynting vector $\boldsymbol{S}(\boldsymbol{r},t) = \mathrm{Re}(\boldsymbol{E}) \times \mathrm{Re}(\boldsymbol{H}) = (E_{xo}^2/Z_o)\cos^2(k_o z - \omega t)\,\hat{\boldsymbol{z}}$ over the cross-sectional area $A$ and duration $\tau$ of the light pulse — recalling that $\varepsilon_o c = \sqrt{\varepsilon_o/\mu_o} = 1/Z_o$. Since the EM momentum-density is known to be $\boldsymbol{p}_{\text{EM}} = \boldsymbol{S}/c^2$, the total EM momentum of the packet is readily found to be[3,4]

$$\boldsymbol{p}_{\text{EM}} = \iiint [\boldsymbol{S}(\boldsymbol{r},t)/c^2]\mathrm{d}x\mathrm{d}y\mathrm{d}z = \tfrac{1}{2}(E_{xo}^2/Z_o c)A\tau\,\hat{\boldsymbol{z}} = (\mathcal{E}/c)\hat{\boldsymbol{z}}. \tag{3}$$

A light pulse can also carry EM angular momentum (density: $\boldsymbol{\mathcal{L}}_{\text{EM}}(\boldsymbol{r},t) = \boldsymbol{r} \times \boldsymbol{p}_{\text{EM}}(\boldsymbol{r},t)$),[3,4] but deriving the relation between the total energy $\mathcal{E}$ and the overall angular momentum $\boldsymbol{L}_{\text{EM}}$ of the pulse requires a more nuanced accounting for the field distribution near the packet boundaries. If the EM field is circularly polarized, for instance, one can show that $\boldsymbol{L}_{\text{EM}} = \pm(\mathcal{E}/\omega)\hat{\boldsymbol{z}}$, where the $\pm$ signs correspond, respectively, to the states of left- and right-circular polarization. In what follows, we shall explore some of the fundamental properties of the EM linear and angular momenta using a number of well-known (and also not-so-well-known) thought experiments.

**3. Relativistic treatment of a light pulse and its emitter**. Figure 2 shows a small plate of mass $m$, located in the $xy$-plane, emitting a light pulse of energy $\mathcal{E}$ along the $z$-axis and, in the process, recoiling with velocity $\boldsymbol{u}$ in the opposite direction. In the plate's initial rest frame $xyz$, the mass $m$ will be reduced to $\widetilde{m}$, and the conservation laws of energy and linear momentum demand that

$$\gamma_u \widetilde{m} c^2 + \mathcal{E} = mc^2 \qquad \text{and} \qquad \gamma_u \widetilde{m} \boldsymbol{u} = -(\mathcal{E}/c)\hat{\boldsymbol{z}}, \tag{4}$$

where, $\gamma_u = 1/\sqrt{1-(u/c)^2}$ is the Lorentz-FitzGerald contraction factor.[3] The above equations are readily solved to yield

$$\widetilde{m}/m = \sqrt{1-(2\mathcal{E}/mc^2)} \qquad \text{and} \qquad \boldsymbol{u} = -(\mathcal{E}/c)\hat{\boldsymbol{z}}/[m-(\mathcal{E}/c^2)]. \tag{5}$$

Note that the center of mass/energy of the system remains fixed at $z = 0$ at all times, since the mass $\gamma_u \widetilde{m} = m - (\mathcal{E}/c^2)$ of the emitter now moves with the constant velocity $u$ to the left, while the mass $\mathcal{E}/c^2$ of the light pulse moves at constant velocity $c$ to the right.

Seen from a different reference frame, say, one within which the $xyz$ frame moves to the right at constant velocity $\boldsymbol{v} = v\hat{\boldsymbol{z}}$, the momentum-energy 4-vector $(p_x, p_y, p_z, \mathcal{E}/c) = (0, 0, \mathcal{E}/c, \mathcal{E}/c)$ of the light pulse becomes[3-5]

$$p'_x = p'_y = 0, \quad p'_z = \gamma_v\left(p_z + \frac{v\mathcal{E}}{c^2}\right) = \sqrt{\frac{1+(v/c)}{1-(v/c)}}\left(\frac{\mathcal{E}}{c}\right). \tag{6a}$$

$$\mathcal{E}' = \gamma_v(\mathcal{E} + vp_z) = \sqrt{\frac{1+(v/c)}{1-(v/c)}}\,\mathcal{E}. \tag{6b}$$

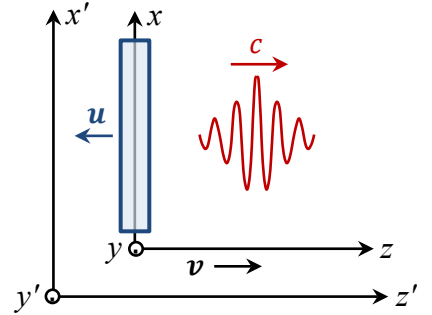

**Fig.2**. In the rest-frame $xyz$ of an emitter of mass $m$ located in the $xy$-plane, a light pulse of energy $\mathcal{E}$ leaves the emitter and travels with velocity $c$ along the z-axis. The emitter's recoil velocity is denoted by $\boldsymbol{u}$. The same event is also seen by a second observer in the reference frame $x'y'z'$, within which $xyz$ travels with the constant velocity $\boldsymbol{v}$ along the positive z-axis.

It is seen that the light is now Doppler-shifted (since energy is proportional to the frequency $\omega$), but its energy and momentum continue to be related via $\boldsymbol{p}' = (\mathcal{E}'/c)\hat{\boldsymbol{z}}$. The energy and momentum of the emitter can be similarly determined in the $x'y'z'$ frame, as follows:

$$\mathcal{E}'_{\text{emitter}} = \gamma_v[\gamma_u\widetilde{m}c^2 + v(\gamma_u\widetilde{m}u)] = \gamma_v[(mc^2 - \mathcal{E}) - (\mathcal{E}v/c)] = \gamma_v mc^2 - \sqrt{\frac{1+(v/c)}{1-(v/c)}}\,\mathcal{E}. \tag{7a}$$

$$p'_{z(\text{emitter})} = \gamma_v[\gamma_u\widetilde{m}u + (v/c^2)\gamma_u\widetilde{m}c^2] = \gamma_v\{[m - (\mathcal{E}/c^2)]v - (\mathcal{E}/c)\} = \gamma_v mv - \sqrt{\frac{1+(v/c)}{1-(v/c)}}\left(\frac{\mathcal{E}}{c}\right). \tag{7b}$$

Clearly, both energy and linear momentum are conserved in the $x'y'z'$ frame—as they were in the $xyz$ frame—and also the center of mass/energy of the system as seen in the $x'y'z'$ frame is moving at constant velocity $\boldsymbol{v}$ along the positive z-axis.

If the emitted light pulse happens to have angular momentum $\boldsymbol{L} = L\hat{\boldsymbol{z}}$, then the emitter will acquire the same angular momentum in the opposite direction, that is, $\boldsymbol{L}_{\text{emitter}} = -L\hat{\boldsymbol{z}}$. Considering that the angular momentum of the emitter remains the same when observed from the $x'y'z'$ frame, we conclude that the angular momentum of the light pulse seen in the $x'y'z'$ frame cannot depend on its frequency $\omega$. This is because the Doppler shift modifies the emitted frequency in different reference frames that move relative to each other at constant velocity along the z-axis, whereas the angular momentum of the pulse must remain the same in all these (inertial) reference frames.

**4. Reflection from an ideal mirror**. Let a plane electromagnetic wave shine on a perfectly electrically conducting mirror at normal incidence, as shown in Fig.3(a). The incident $E$ and $H$ fields are given by

$$\boldsymbol{E}(\boldsymbol{r}, t) = E_o\hat{\boldsymbol{x}}\exp[\mathrm{i}(k_o z - \omega t)]. \tag{8a}$$

$$\boldsymbol{H}(\boldsymbol{r}, t) = (E_o/Z_o)\hat{\boldsymbol{y}}\exp[\mathrm{i}(k_o z - \omega t)]. \tag{8b}$$

The incident $E$-field excites the electric current-density $\boldsymbol{J}(\boldsymbol{r}, t) = 2(E_o/Z_o)\exp(-\mathrm{i}\omega t)\,\delta(z)\hat{\boldsymbol{x}}$ on the front facet of the mirror, which radiates the following plane-waves to the right and left of the mirror:

$$\boldsymbol{E}^{(+)}(\boldsymbol{r}, t) = -E_o\hat{\boldsymbol{x}}\exp[\mathrm{i}(k_o z - \omega t)], \qquad z \geq 0. \tag{9}$$

$$\boldsymbol{E}^{(-)}(\boldsymbol{r}, t) = -E_o\hat{\boldsymbol{x}}\exp[-\mathrm{i}(k_o z + \omega t)], \qquad z \leq 0. \tag{10}$$

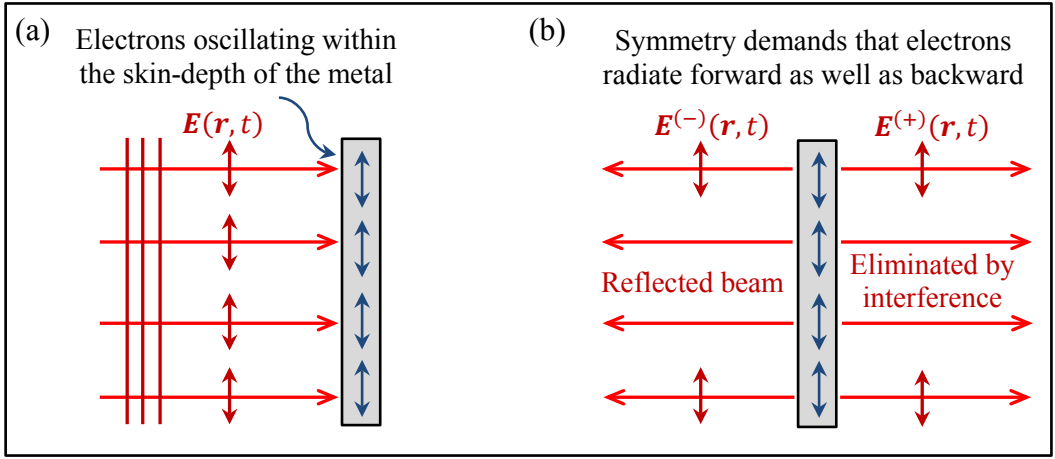

**Fig.3**. (a) A linearly-polarized plane-wave excites the conduction electrons at the front facet of a perfect electrically conducting mirror. (b) The excited electrons radiate a pair of plane-waves into the surrounding free space. Whereas the left-propagating radiated wave constitutes the reflected beam from the mirror, the right-propagating radiated wave is cancelled by destructive interference with the continuation of the incident beam beyond the mirror.

On the right-hand side, the continuation of the incident beam cancels the effects of $\boldsymbol{E}^{(+)}(\boldsymbol{r},t)$, whereas, on the left-hand side, the left-propagating $\boldsymbol{E}^{(-)}(\boldsymbol{r},t)$ gives rise to the reflected beam from the mirror. The net (linear) momentum carried by the pair of right- and left-propagating plane-waves is, of course, equal to zero. However, the incident $B$-field exerts a force on the surface current $\boldsymbol{J}(\boldsymbol{r},t)$ that is the well-known radiation pressure (i.e., time-averaged force per unit area) on the mirror, as follows:

$$\text{Radiation pressure} = \langle \int_{-\infty}^{\infty} \text{Re}[\boldsymbol{J}(\boldsymbol{r},t)] \times \text{Re}[\boldsymbol{B}(\boldsymbol{r},t)] \mathrm{d}z \rangle = (2E_\circ/Z_\circ)\hat{\boldsymbol{x}} \times \mu_\circ(E_\circ/Z_\circ)\hat{\boldsymbol{y}} \langle \cos^2(\omega t) \rangle$$

$$= (E_0^2/Z_\circ)\hat{\boldsymbol{z}}/c = 2\langle \boldsymbol{S}(\boldsymbol{r},t)\rangle/c. \tag{11}$$

In the above equation, $\langle \boldsymbol{S}(\boldsymbol{r},t)\rangle = E_\circ\hat{\boldsymbol{x}} \times H_\circ\hat{\boldsymbol{y}} \langle \cos^2(k_\circ z - \omega t)\rangle = \tfrac{1}{2}(E_0^2/Z_\circ)\hat{\boldsymbol{z}}$ is the time-averaged Poynting vector of the incident plane-wave. Considering that the speed of light in vacuum is $c$, and that the radiation pressure on the mirror must equal twice the incident momentum per unit area per unit time, we conclude that the momentum density of the incident wave is $\boldsymbol{p}_{\text{EM}}(\boldsymbol{r},t) = \langle \boldsymbol{S}(\boldsymbol{r},t)\rangle/c^2$.

In a variation on the thought experiment depicted in Fig.2, we now allow the emitted pulse to bounce back from a perfect electrically conducting mirror of mass $m$ located at $z = \ell$, return to the emitter, and be fully absorbed at the emitter. Conservation of energy and linear momentum within the $xyz$ frame, for the system consisting of the light-pulse and the reflector, yield

$$\tilde{\mathcal{E}} = \mathcal{E}_{\text{pulse}}^{(\text{reflected})} = \frac{\mathcal{E}}{1+2(\mathcal{E}/mc^2)}. \tag{12a}$$

$$\mathcal{E}_{\text{reflector}} = \gamma_u mc^2 = mc^2 + \frac{2\mathcal{E}^2}{mc^2 + 2\mathcal{E}}. \tag{12b}$$

$$\boldsymbol{p}_{\text{reflector}} = \gamma_u m\boldsymbol{u} = 2\left[\frac{1+(\mathcal{E}/mc^2)}{1+2(\mathcal{E}/mc^2)}\right](\mathcal{E}/c)\hat{\boldsymbol{z}}. \tag{12c}$$

The reflected wavepacket now returns to the emitter and gets absorbed there, its momentum and energy being added to the momentum and energy of the receding emitter. We will have

$$\mathcal{E}_{\text{emitter}} = (mc^2 - \mathcal{E}) + \tilde{\mathcal{E}} = mc^2 - \frac{2\mathcal{E}^2}{mc^2 + 2\mathcal{E}}. \tag{13a}$$

$$\boldsymbol{p}_{\text{emitter}} = -(\mathcal{E}/c)\hat{\boldsymbol{z}} - (\tilde{\mathcal{E}}/c)\hat{\boldsymbol{z}} = -2\left[\frac{1 + (\mathcal{E}/mc^2)}{1 + 2(\mathcal{E}/mc^2)}\right](\mathcal{E}/c)\hat{\boldsymbol{z}}. \tag{13b}$$

Clearly, the overall energy of the system (i.e., emitter plus reflector) is constant at $2mc^2$, and the overall momentum of the system is zero, indicating that the center of mass/energy is *not* shifting with the passage of time. It is not difficult to confirm that the center of mass/energy of the system remains at $z = \ell/2$ at all times.

In the case of a light pulse with angular momentum $L_{\text{EM}}\hat{\boldsymbol{z}}$ propagating forward and then backward between an emitter/absorber (located at $z = 0$) and a perfect mirror (located at $z = \ell$), the emitter starts to rotate at a constant angular velocity around the $z$-axis as soon as the pulse leaves the emitter. The pulse maintains its angular momentum even after being reflected from the mirror at $z = \ell$, since the forward-radiated beam must cancel out the incident beam—recall that the forward-radiated beam has a $\pi$ phase-shift relative to the incident beam. Due to symmetry, the backward-radiated beam must have the same angular momentum as the forward-radiated beam. Consequently, no angular momentum is transferred from the light pulse to the reflector and, therefore, the net torque acting on the reflector will be zero. Upon arrival at the emitter, the light pulse will get absorbed, returning its angular momentum back to the emitter. The emitter thus stops rotating, albeit with its angular position now changed relative to its initial orientation. The system is seen to undergo an overall twist (around the $z$-axis) due to its internal processes, but a twist is very different from a displacement of the center of mass/energy. In this thought experiment, there is no useful information in the twisted position of the emitter/absorber that one could exploit—along the lines of Einstein's argument presented in Sec.6—in order to relate the angular momentum $\boldsymbol{L}_{\text{EM}}$ of the light pulse to its energy $\mathcal{E}$. All one can say is that, for a given number of photons, the EM angular momentum must be independent of the oscillation frequency $\omega$ of the wavepacket, as was pointed out in the last paragraph of Sec.3.

**5. Angular momentum radiated by a current-carrying plate**. With reference to Fig.4, let a thin current-carrying sheet located in the $xy$-plane at $z = 0$ radiate a pair of circularly-polarized, zeroth-order Bessel beams into its surrounding region. The Bessel beam propagating to the right (along $+z$) is a superposition of transverse electric (TE) and transverse magnetic (TM) modes, whose $E$ and $H$ fields are given by [4,5,9,10]

$$\boldsymbol{E}^{(+)}(\boldsymbol{r}, t) = E_{z0}[(i\sigma_z/\sigma_r)J_1(k_o\sigma_r r)\hat{\boldsymbol{r}} - (1/\sigma_r)J_1(k_o\sigma_r r)\hat{\boldsymbol{\varphi}} - J_0(k_o\sigma_r r)\hat{\boldsymbol{z}}]\exp[i(k_o\sigma_z z - \omega t)]. \tag{14a}$$

$$\boldsymbol{H}^{(+)}(\boldsymbol{r}, t) = (E_{z0}/Z_0)[(\sigma_z/\sigma_r)J_1(k_o\sigma_r r)\hat{\boldsymbol{r}} + (i/\sigma_r)J_1(k_o\sigma_r r)\hat{\boldsymbol{\varphi}} + iJ_0(k_o\sigma_r r)\hat{\boldsymbol{z}}]\exp[i(k_o\sigma_z z - \omega t)]. \tag{14b}$$

In the above equations, $\omega$ is the frequency of the monochromatic beam, $k_o = \omega/c$ is the wavenumber, $\sigma_r$ and $\sigma_z$ are components of a unit vector along the radial and axial directions (with $\sigma_r^2 + \sigma_z^2 = 1$), $E_{z0}$ is the $E$-field amplitude, and $J_0(\cdot)$ and $J_1(\cdot)$ are Bessel functions of the first kind, 0th and 1st order, respectively. The 90° relative phase imposed between the TE and TM modes in Eqs.(14) is responsible for the state of circular polarization of the beam.

The Bessel beam propagating to the left (along $-z$) is similar to the one given above, except for a replacement of $\sigma_z$ with $-\sigma_z$, and a sign-flipping of its TM components (i.e., $E_r, E_z, H_\varphi$), that is,

$$\boldsymbol{E}^{(-)}(\boldsymbol{r}, t) = E_{z0}[(i\sigma_z/\sigma_r)J_1(k_o\sigma_r r)\hat{\boldsymbol{r}} - (1/\sigma_r)J_1(k_o\sigma_r r)\hat{\boldsymbol{\varphi}} + J_0(k_o\sigma_r r)\hat{\boldsymbol{z}}]\exp[-i(k_o\sigma_z z + \omega t)]. \tag{15a}$$

$$\boldsymbol{H}^{(-)}(\boldsymbol{r}, t) = (E_{z0}/Z_0)[-(\sigma_z/\sigma_r)J_1(k_o\sigma_r r)\hat{\boldsymbol{r}} - (i/\sigma_r)J_1(k_o\sigma_r r)\hat{\boldsymbol{\varphi}} + iJ_0(k_o\sigma_r r)\hat{\boldsymbol{z}}]\exp[-i(k_o\sigma_z z + \omega t)]. \tag{15b}$$

Note that the light emerging on the right-hand side of the radiator is left-circularly-polarized (LCP), whereas that emerging on the left-hand side is right-circularly-polarized (RCP), resulting in their exerted torques on the radiating sheet being in the same direction and, therefore, additive. Aside from the spin angular momentum associated with their circular polarization states, each beam also carries a $2\pi$ vortex (with an accompanying orbital angular momentum), as the phase of the circular polarization varies continuously in a roundtrip around the $z$-axis. The spin and orbital angular momenta of each beam are thus equal in magnitude and opposite in orientation, so that the net angular momentum carried by each beam equals zero.

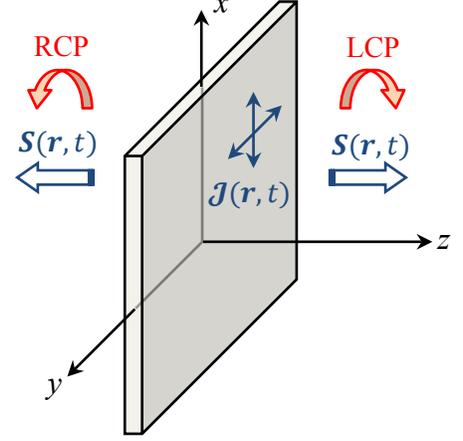

**Fig.4**. A thin sheet carrying the electric current-density $\boldsymbol{J}(\boldsymbol{r},t)$ is located in the $xy$-plane at $z = 0$. The current along the $y$-axis is phase-shifted by 90° relative to that along the $x$-axis, resulting in a pair of circularly-polarized light beams emerging on the right and left sides of the emitter.

The Poynting vector $\boldsymbol{S}^{(\pm)}(\boldsymbol{r},t)$ on each side of the radiating sheet may now be computed, as follows:

$$\boldsymbol{S}^{(\pm)}(\boldsymbol{r},t) = \text{Re}[\boldsymbol{E}^{(\pm)}] \times \text{Re}[\boldsymbol{H}^{(\pm)}]$$
$$= (\sigma_z E_{z0}^2/Z_0)[-(1/\sigma_r)J_0(k_0\sigma_r r)J_1(k_0\sigma_r r)\hat{\boldsymbol{\varphi}} \pm (1/\sigma_r^2)J_1^2(k_0\sigma_r r)\hat{\boldsymbol{z}}]. \qquad (16)$$

Note in Eq.(16) that the azimuthal component $S_\varphi$ of the Poynting vector oscillates between positive and negative values as the radial coordinate $r$ goes from zero to infinity. This is a consequence of the fact that, for each beam, the spin and orbital angular momenta are equal in magnitude and opposite in direction, so that, upon integration over any cross-section of the beam, they cancel each other out.

The discontinuity of the tangential $H$-fields of these counter-propagating Bessel beams at the surfaces $z = 0^\pm$ of the current sheet determines the electric current-density $\boldsymbol{J}(\boldsymbol{r},t)$ within the sheet, as follows:

$$\boldsymbol{J}(\boldsymbol{r},t) = 2(E_{z0}/Z_0)[-(i/\sigma_r)J_1(k_0\sigma_r r)\hat{\boldsymbol{r}} + (\sigma_z/\sigma_r)J_1(k_0\sigma_r r)\hat{\boldsymbol{\varphi}}]\delta(z)\exp(-i\omega t). \qquad (17)$$

Similarly, the electric charge-density $\rho(\boldsymbol{r},t)$ within the sheet can be determined from the discontinuity at $z = 0$ of $E_z$, the perpendicular component of the $E$-field, that is,

$$\rho(\boldsymbol{r},t) = -2\varepsilon_0 E_{z0} J_0(k_0\sigma_r r)\delta(z)\exp(-i\omega t). \qquad (18)$$

**Digression**: Equations (17) and (18) satisfy the charge-current continuity equation,[4] namely,

$\boldsymbol{\nabla}\cdot\boldsymbol{J} + \partial\rho/\partial t = r^{-1}\partial(rJ_r)/\partial r + r^{-1}\partial J_\varphi/\partial\varphi + \partial\rho/\partial t$

$= \{-2(E_{z0}/Z_0)(i/\sigma_r)[r^{-1}J_1(k_0\sigma_r r) + k_0\sigma_r J_0(k_0\sigma_r r) - r^{-1}J_1(k_0\sigma_r r)] + 2i\omega\varepsilon_0 E_0 J_0(k_0\sigma_r r)\}\delta(z)\exp(-i\omega t)$

$= -2iE_{z0}[(k_0/Z_0) - \varepsilon_0\omega]J_0(k_0\sigma_r r)\delta(z)\exp(-i\omega t) = 0.$

We are now in a position to compute the time-rate of radiation of EM energy per unit area of the current sheet from the knowledge of the current-density $\boldsymbol{J}(\boldsymbol{r},t)$ given by Eq.(17), as well as the tangential $E$-field components given by Eqs.(14a) and (15a). We find

$$\dot{\mathcal{E}}(\boldsymbol{r},t) = -\text{Re}(\boldsymbol{E})\cdot\text{Re}(\boldsymbol{J}) = 2(E_{z0}^2/Z_0)(\sigma_z/\sigma_r^2)J_1^2(k_0\sigma_r r)[\sin^2(\omega t) + \cos^2(\omega t)]\delta(z). \qquad (19)$$

Integrating Eq.(16) over the $xy$-plane of the sheet within a disk of large radius $R$ yields

$$\int_{r=0}^{R}\int_{z=-\infty}^{\infty} 2\pi r \dot{\mathcal{E}}(\boldsymbol{r},t)\mathrm{d}r\mathrm{d}z = 4\pi(E_{z0}^2/Z_0)(\sigma_z/\sigma_r^2)\int_0^R r J_1^2(k_o\sigma_r r)\mathrm{d}r$$

Gradshteyn & Ryzhik[13] 5.54-2 $\rightarrow = 2\pi(E_{z0}^2/Z_0)(\sigma_z/\sigma_r^2)R^2[J_1^2(k_o\sigma_r R) - J_0(k_o\sigma_r R)J_2(k_o\sigma_r R)]$

$$\cong 4E_{z0}^2 \sigma_z R/(Z_0 k_o \sigma_r^3); \qquad (k_o\sigma_r R \gg 1). \qquad (20)$$

$J_n(x) \to \sqrt{2/(\pi x)} \cos[x - (n\pi/2) - (\pi/4)]$

Next, we compute the torque per unit area exerted by the radiated EM field of Eqs.(14) and (15) on the current-density of Eq.(17) and charge-density of Eq.(18), as follows:

$$\boldsymbol{\tau}(\boldsymbol{r},t) = \boldsymbol{r} \times [\mathrm{Re}(\rho)\mathrm{Re}(\boldsymbol{E}) + \mathrm{Re}(\boldsymbol{\mathcal{J}}) \times \mathrm{Re}(\mu_o \boldsymbol{H})] = r[\mathrm{Re}(\rho)\mathrm{Re}(E_\varphi) - \mu_o \mathrm{Re}(J_r)\mathrm{Re}(H_z)]\hat{\boldsymbol{z}}$$

$$= (2\varepsilon_o E_{z0}^2/\sigma_r) r J_0(k_o\sigma_r r) J_1(k_o\sigma_r r)[\cos^2(\omega t) + \sin^2(\omega t)]\delta(z)\hat{\boldsymbol{z}}. \qquad (21)$$

Integration over the $xy$-plane of the sheet within a disk of large radius $R$ now yields the following expression for the total torque exerted on the current-carrying sheet:

$$\int_{r=0}^{R}\int_{z=-\infty}^{\infty} 2\pi r \tau_z(\boldsymbol{r},t)\mathrm{d}r\mathrm{d}z = (4\pi\varepsilon_o E_{z0}^2/\sigma_r)\int_0^R r^2 J_0(k_o\sigma_r r)J_1(k_o\sigma_r r)\mathrm{d}r$$

$$= (4\pi\varepsilon_o E_{z0}^2/k_o^3\sigma_r^4)\int_0^{k_o\sigma_r R} x^2 J_0(x)J_1(x)\mathrm{d}x$$

$J_1(x) = -J_0'(x)$; integration by parts $\rightarrow = (4\pi\varepsilon_o E_{z0}^2/k_o^3\sigma_r^4)\left[-\tfrac{1}{2}x^2 J_0^2(x)\big|_{x=0}^{k_o\sigma_r R} + \int_0^{k_o\sigma_r R} x J_0^2(x)\mathrm{d}x\right]$

Gradshteyn & Ryzhik[13] 5.54-2 $\rightarrow = 2\pi\varepsilon_o E_{z0}^2 R^2 J_1^2(k_o\sigma_r R)/(k_o\sigma_r^2)$

$J_1(x) \to \sqrt{2/(\pi x)} \sin(x - \tfrac{1}{4}\pi)$ $\rightarrow \cong 4\varepsilon_o E_{z0}^2 R \sin^2(k_o\sigma_r R - \tfrac{1}{4}\pi)/(k_o^2\sigma_r^3)$

$$= 4E_{z0}^2 R \sin^2(k_o\sigma_r R - \tfrac{1}{4}\pi)/(Z_0 k_o\sigma_r^3 \omega); \quad (k_o\sigma_r R \gg 1). \quad (22)$$

The sinusoidal variation of the total torque as a function of $R$ indicates that the radiated spin and orbital momenta have differing strengths in different cross-sectional regions of each beam, so that integration over $r$ from 0 to $R$ leaves some residual torque in between the various zeros of the Bessel function $J_1(\cdot)$. Aside from the sinusoidal factor in Eq.(22), it is seen that the radiated energy of Eq.(20) and the (residual) radiated angular momentum of Eq.(22) are in the ratio of $\sigma_z \omega : 1$. Recalling that, in quantum-optical language, each radiated photon of energy $\hbar\omega$ carries a spin angular momentum of $\pm\hbar$ and also an orbital angular momentum of $\mp\hbar$ (assuming $2\pi$ vorticity), it should not be surprising that the ratio of radiated energy to the residual angular momentum in the present example is proportional to $\omega$.

**6. The Einstein-box gedanken experiment.** In the original Einstein-box thought experiment,[3,5] the linear momentum of an electromagnetic wavepacket of energy $\mathcal{E}$ propagating along the $z$-axis was shown to be $\boldsymbol{p}_{\mathrm{EM}} = (\mathcal{E}/c)\hat{\boldsymbol{z}}$. Here, we present a slightly modified version of the original thought experiment that avoids complications that might arise from perceived violations of special relativity.

With reference to Fig.5, consider an emitter and an absorber of EM radiation sitting on a frictionless rail and facing each other. Initially both emitter and absorber are stationary, their masses are $M_e$ and $M_a$, and their separation is $L$. At $t = 0$, a short EM pulse of energy $\mathcal{E}$ and momentum $\boldsymbol{p}_{\mathrm{EM}}$ is emitted toward the absorber. The backlash causes the emitter to move backward with a constant velocity $V_e = p_{\mathrm{EM}}/(M_e - \mathcal{E}/c^2)$. At $t = L/c$, the pulse arrives at the absorber and is fully absorbed. The absorber thus acquires the momentum $\boldsymbol{p}_{\mathrm{EM}}$ and moves forward with a constant velocity $V_a = p_{\mathrm{EM}}/(M_a + \mathcal{E}/c^2)$. At any time $t > L/c$, the center-of-mass of the system will be at

$$z_{CM}(t) = \{(M_a + \mathcal{E}/c^2)[L + V_a(t - L/c)] - (M_e - \mathcal{E}/c^2)V_e t\}/(M_e + M_a)$$

$$= [(M_a + \mathcal{E}/c^2)L + p_{EM}(t - L/c) - p_{EM}t]/(M_e + M_a)$$

$$= [M_a + (\mathcal{E}/c^2) - (p_{EM}/c)]L/(M_e + M_a); \qquad t > L/c. \qquad (23)$$

However, this center-of-mass/energy cannot have moved from its initial position at $z_{CM}(0) = M_a L/(M_e + M_a)$, as there are no external forces acting on the system. Therefore,

$$z_{CM}(t) = z_{CM}(0) \quad \rightarrow \quad M_a + (\mathcal{E}/c^2) - (p_{EM}/c) = M_a \quad \rightarrow \quad p_{EM} = \mathcal{E}/c. \qquad (24)$$

We conclude that the EM momentum of the wavepacket of energy $\mathcal{E}$ in free space is $\boldsymbol{p}_{EM} = (\mathcal{E}/c)\hat{\boldsymbol{z}}$.

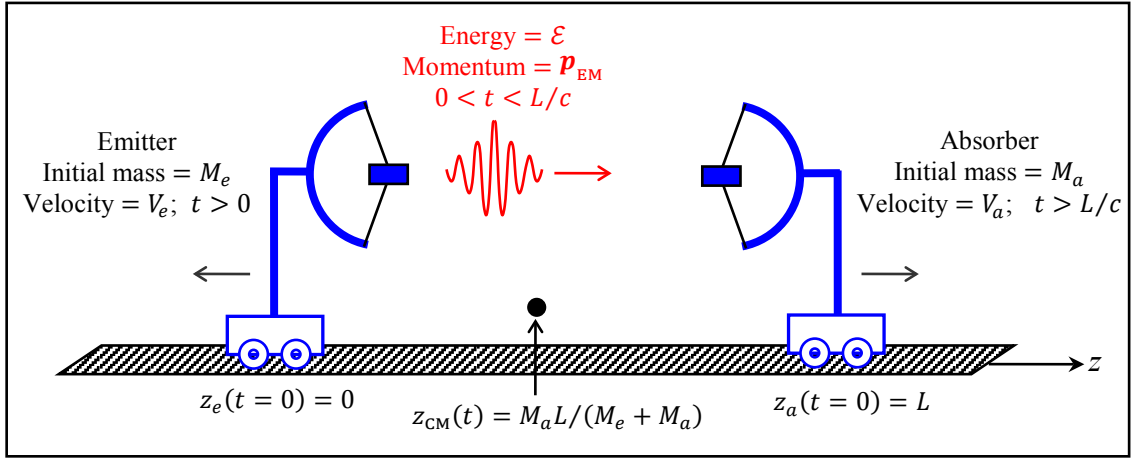

**Fig. 5**. An emitter of EM waves rests on a frictionless rail at $z_e = 0$. At $t = 0$, the emitter sends a short pulse of energy $\mathcal{E}$ and unknown momentum $\boldsymbol{p}_{EM}$ toward a distant receiver. The resulting backlash imparts a momentum $-\boldsymbol{p}_{EM}$ to the emitter, causing it to move to the left with a constant velocity $V_e$. Denoting the initial mass of the emitter by $M_e$, we find $\boldsymbol{V}_e = -\boldsymbol{p}_{EM}\hat{\boldsymbol{z}}/(M_e - \mathcal{E}/c^2)$. At $t = L/c$, the pulse arrives at the stationary absorber of mass $M_a$ located at $z_a = L$. Upon absorption, the pulse transfers its entire energy and momentum to the absorber, which subsequently moves to the right at a constant velocity $\boldsymbol{V}_a = \boldsymbol{p}_{EM}\hat{\boldsymbol{z}}/(M_a + \mathcal{E}/c^2)$. Since no external forces act on the system, its center of mass/energy must remain at $z_{CM} = M_a L/(M_e + M_a)$ at all times.

**7. The thought experiment of Balazs.**[11,12] Balazs's variant of the Einstein-box experiment, depicted in Fig.6, features a short pulse of light and a transparent slab of length $L$ and mass $M$. In the free-space region outside the slab, the pulse, having energy $\mathcal{E} = mc^2$ and momentum $\boldsymbol{p}_{EM} = mc\hat{\boldsymbol{z}}$, travels with the speed $c$ of light in vacuum. Inside the slab, the pulse travels with the group velocity $V_g$, which is generally less than $c$. The entrance and exit facets of the slab are anti-reflection coated to ensure the passage of the entire pulse through the slab.

In one experiment, the pulse travels entirely in the free-space region outside the slab, while in another, the pulse spends a fraction of its time inside the slab. In the latter case, upon re-emergence into free space, the pulse is seen to have fallen behind by a distance of $[(c/V_g) - 1]L$ relative to the position it would have had (along the $z$-axis), had it remained in free space the entire time. Since no external forces are at work here, the center-of-mass of the system (consisting of the light pulse and the slab) at any given instant of time must be in the same location in the two experiments. Requiring the center-of-mass/energy of the system to be at the expected location at the end of the second experiment thus demands that the slab be displaced forward by a distance $\Delta z$ such that

$$M\Delta z = (\mathcal{E}/c^2)[(c/V_g) - 1]L. \qquad (25)$$

During the time interval $L/V_g$ when the pulse is inside the (transparent) slab, given that the slab must move forward by $\Delta z$ during this time interval, the slab maintains a velocity $\Delta z/(L/V_g)$ and, therefore, a momentum $M\Delta z(V_g/L) = (\mathcal{E}/c)[1 - (V_g/c)]$. Recognizing that the total available momentum is the initial momentum of the light pulse, namely, $p_{EM} = \mathcal{E}/c$, we conclude that the EM momentum of the pulse, while inside the slab, must have been reduced to $(\mathcal{E}/c)(V_g/c)$.

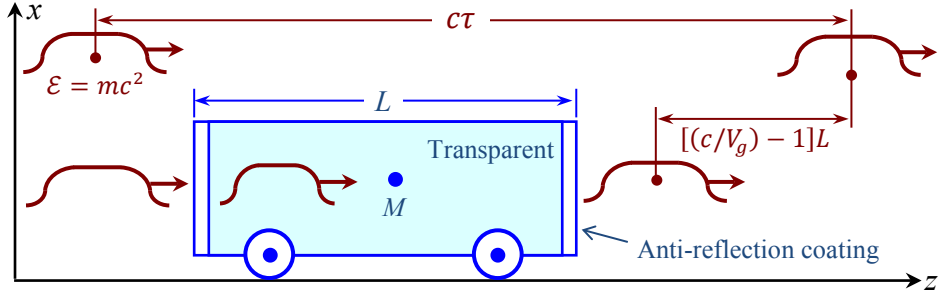

**Fig.6**. Balazs's variant of the Einstein box experiment features a short pulse of light and a massive, transparent slab of length $L$ and mass $M$. In the free-space region outside the slab, the pulse, having energy $\mathcal{E} = mc^2$ and momentum $\boldsymbol{p}_{EM} = mc\hat{\boldsymbol{z}}$, travels with speed $c$. Inside the slab, the pulse travels with the group velocity $V_g$. The entrance and exit facets of the slab are anti-reflection coated to ensure the passage of the entire pulse through the slab. In one experiment, the pulse travels entirely in the free-space region outside the slab, while in another, the pulse spends a fraction of its time inside the slab. Since no external forces are at work, the center of mass of the system (i.e., light pulse + slab) at any given moment must be in exactly the same location in the two experiments.

Neither the light beam's cross-sectional area nor its duration could be any different inside and outside the slab. Consequently, for the EM energy to remain unchanged, the Poynting vector $\boldsymbol{S}$ must remain the same inside and outside. When inside the slab, however, the length of the pulse along the $z$-axis shrinks by a factor of $V_g/c$, as its propagation velocity drops from $c$ to $V_g$. The reduction of the spatial volume occupied by the pulse by the factor $V_g/c$ would entirely account for its reduced momentum by the same factor if the EM momentum density inside the slab remains at $\boldsymbol{p}_{EM} = \boldsymbol{S}/c^2$. The corresponding EM momentum is known as the Abraham momentum which, following Balazs' argument, constitutes the EM part of the momentum of the light pulse inside the host medium.

**7.1. Energy content of a light pulse inside a homogeneous and transparent dielectric slab**. For a plane-wave of frequency $\omega$, propagating along the $z$-axis inside a homogeneous medium of refractive index $n(\omega)$, as depicted in Fig.7, the propagation vector is $\boldsymbol{k}(\omega) = [n(\omega)\omega/c]\hat{\boldsymbol{z}}$, and the $E$-field amplitude is $\boldsymbol{E}(\omega) = [E_x(\omega)\hat{\boldsymbol{x}} + E_y(\omega)\hat{\boldsymbol{y}}]\exp[\mathrm{i}(\boldsymbol{k}\cdot\boldsymbol{r} - \omega t)]$. According to the Maxwell equation $\boldsymbol{\nabla}\times\boldsymbol{E} = -\partial_t\boldsymbol{B}$, one will have $\boldsymbol{k}(\omega)\times\boldsymbol{E}(\omega) = \mu_0\omega\boldsymbol{H}(\omega)$. Recalling that the speed of light in vacuum is $c = (\mu_0\varepsilon_0)^{-\frac{1}{2}}$ and that the impedance of free space is $Z_0 = (\mu_0/\varepsilon_0)^{\frac{1}{2}}$, we write

$$\boldsymbol{E}(\boldsymbol{r},t) = (2\pi)^{-1}\int_{-\infty}^{\infty} \boldsymbol{E}(\omega)\exp[\mathrm{i}(\boldsymbol{k}\cdot\boldsymbol{r} - \omega t)]\,\mathrm{d}\omega, \tag{26a}$$

$$\boldsymbol{H}(\boldsymbol{r},t) = (2\pi Z_0)^{-1}\int_{-\infty}^{\infty} n(\omega)\hat{\boldsymbol{z}}\times\boldsymbol{E}(\omega)\exp[\mathrm{i}(\boldsymbol{k}\cdot\boldsymbol{r} - \omega t)]\,\mathrm{d}\omega. \tag{26b}$$

In these equations, $\boldsymbol{E}(\omega)$ is a Hermitian function of $\omega$, that is, $\boldsymbol{E}(-\omega) = \boldsymbol{E}^*(\omega)$, which is required by the fact that $\boldsymbol{E}(\boldsymbol{r},t)$ and $\boldsymbol{H}(\boldsymbol{r},t)$ are real-valued functions of the spacetime. Moreover, $n(\omega)$ is Hermitian. If we confine our attention to a fairly narrow range of frequencies $\omega$, centered around $\pm\omega_0$, within which $\boldsymbol{E}(\omega) \neq 0$, it is possible to assume that $n(\omega)$ is real-valued as well, and that, therefore, the host medium is transparent.

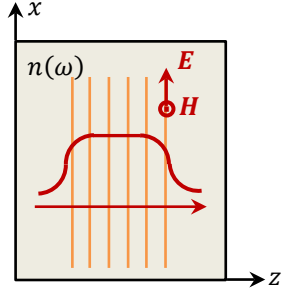

**Fig.7**. Within a homogeneous, isotropic, linear, transparent medium of refractive index $n(\omega)$, a light pulse having a large cross-sectional area and center frequency $\omega_0$ propagates along the $z$-axis. The beam is linearly polarized, with its electric field $\boldsymbol{E}$ aligned with the $x$-axis and its magnetic field $\boldsymbol{H}$ aligned with the $y$-axis. The pulse propagates at the group velocity $v_g = c/n_g$, where $n_g = \mathrm{d}[\omega n(\omega)]/\mathrm{d}\omega|_{\omega_0}$.

The rate of flow of energy within the medium of refractive index $n(\omega)$ is given by the component of the Poynting vector $\boldsymbol{S}(\boldsymbol{r},t) = \boldsymbol{E}(\boldsymbol{r},t) \times \boldsymbol{H}(\boldsymbol{r},t)$ along the propagation direction $\hat{\boldsymbol{z}}$. We are assuming here a uniform EM field amplitude in the $xy$-plane. The integral of $S_z(\boldsymbol{r}_0,t)$ over the time $t$ thus yields the energy of the EM beam per unit cross-sectional area, as follows:

$$\mathcal{E}(\boldsymbol{r}_0) = \int_{-\infty}^{\infty} S_z(\boldsymbol{r}_0,t)\mathrm{d}t$$

$$= (4\pi^2 Z_0)^{-1} \iiint_{-\infty}^{\infty} n(\omega')\boldsymbol{E}(\omega) \cdot \boldsymbol{E}(\omega') \exp\{\mathrm{i}[\boldsymbol{k}(\omega) + \boldsymbol{k}(\omega')] \cdot \boldsymbol{r}_0\} \exp[-\mathrm{i}(\omega+\omega')t]\,\mathrm{d}\omega\mathrm{d}\omega'\mathrm{d}t$$

$$= (2\pi Z_0)^{-1} \iint_{-\infty}^{\infty} n(\omega')\boldsymbol{E}(\omega) \cdot \boldsymbol{E}(\omega') \exp\{\mathrm{i}[\boldsymbol{k}(\omega) + \boldsymbol{k}(\omega')] \cdot \boldsymbol{r}_0\} \delta(\omega+\omega')\,\mathrm{d}\omega\mathrm{d}\omega'$$

$$= (2\pi Z_0)^{-1} \int_{-\infty}^{\infty} n(\omega)\boldsymbol{E}(\omega) \cdot \boldsymbol{E}^*(\omega)\mathrm{d}\omega. \tag{27}$$

In accordance with Eq.(27), the energy per unit cross-sectional area of the beam can be evaluated at any point $\boldsymbol{r}_0$ inside the transparent medium, with the result $\mathcal{E}(\boldsymbol{r}_0)$ being a real-valued entity that is completely independent of the chosen location $\boldsymbol{r}_0$.

**7.2. Electromagnetic (Abraham) momentum inside the dielectric slab**. To compute the Abraham momentum of the beam inside its transparent host, note that Abraham's linear momentum density is given by $\boldsymbol{p}_A = \boldsymbol{E} \times \boldsymbol{H}/c^2$, and that this volumetric density must be integrated along the entire $z$-axis to yield the EM linear momentum per unit cross-sectional area of the beam. We will have

$$\boldsymbol{p}_A(t) = c^{-2} \int_{-\infty}^{\infty} \boldsymbol{E}(\boldsymbol{r},t) \times \boldsymbol{H}(\boldsymbol{r},t)\mathrm{d}z$$

$$= \tfrac{1}{4\pi^2 c^2 Z_0} \int_{-\infty}^{\infty}\int_{-\infty}^{\infty} \boldsymbol{E}(\omega) \times [n(\omega')\hat{\boldsymbol{z}} \times \boldsymbol{E}(\omega')] \exp[-\mathrm{i}(\omega+\omega')t] \int_{-\infty}^{\infty} \exp\{\mathrm{i}[k(\omega)+k(\omega')]z\}\,\mathrm{d}z\,\mathrm{d}\omega\,\mathrm{d}\omega'$$

$$= \tfrac{\hat{\boldsymbol{z}}}{2\pi c^2 Z_0} \int_{-\infty}^{\infty}\int_{-\infty}^{\infty} n(\omega')\boldsymbol{E}(\omega) \cdot \boldsymbol{E}(\omega') \exp[-\mathrm{i}(\omega+\omega')t]\,\delta[k(\omega)+k(\omega')]\,\mathrm{d}\omega\,\mathrm{d}\omega'$$

$$\cong \tfrac{\hat{\boldsymbol{z}}}{2\pi c^2 Z_0 (\mathrm{d}k/\mathrm{d}\omega)_{\omega_0}} \int_{-\infty}^{\infty} n(-\omega)\boldsymbol{E}(\omega) \cdot \boldsymbol{E}(-\omega) \exp[-\mathrm{i}(\omega-\omega)t]\,\mathrm{d}\omega$$

$$= \tfrac{\hat{\boldsymbol{z}}}{2\pi c Z_0 n_g(\omega_0)} \int_{-\infty}^{\infty} n(\omega)\boldsymbol{E}(\omega) \cdot \boldsymbol{E}^*(\omega)\mathrm{d}\omega = \mathcal{E}\hat{\boldsymbol{z}}/[cn_g(\omega_0)]. \tag{28}$$

In the above derivation, we have invoked the fact that $k(\omega)$ needs to be evaluated only in a narrow range of frequencies around $\pm\omega_0$, where

$$k(\omega) \cong \pm n(\pm\omega_o)\omega_o/c + \tfrac{d}{d\omega}[n(\omega)\omega/c]_{\omega=\pm\omega_o}(\omega \mp \omega_o) + \tfrac{d^2}{2d\omega^2}[n(\omega)\omega/c]_{\omega=\pm\omega_o}(\omega \mp \omega_o)^2. \quad (29)$$

Note that $k(\omega) + k(\omega') = 0$ when $\omega' = -\omega$, and that $d[n(\omega)\omega/c]/d\omega$ is an even function of $\omega$, whereas $d^2[n(\omega)\omega/c]/d\omega^2$ is an odd function. Consequently, in the immediate vicinity of $\omega' = -\omega$, one may write $k(\omega) + k(\omega') \cong d[n(\omega)\omega/c]/d\omega|_{\omega_o}(\omega + \omega')$ to a good approximation. This implies that, when $\omega$ and $\omega'$ hover around $\pm\omega_o$, we may write

$$\delta[k(\omega) + k(\omega')] \cong \frac{c}{d[\omega n(\omega)]/d\omega|_{\omega_o}} \delta(\omega + \omega'). \quad (30)$$

**7.3. Alternative evaluation of the energy content of the beam inside the slab**. In this section, we analyze the energy content of the EM wavepacket shown in Fig.7 from a somewhat different perspective, and arrive at the same result as in Eq.(27). It is well-known that the energy-densities of the electric and magnetic fields are given by $\tfrac{1}{2}\varepsilon_o \boldsymbol{E} \cdot \boldsymbol{E}$ and $\tfrac{1}{2}\mu_o \boldsymbol{H} \cdot \boldsymbol{H}$, respectively. To these we must add the kinetic and potential energy densities of the electric dipoles that are excited by the resident EM field, namely,

$$\mathcal{E}_K(\boldsymbol{r},t) = \tfrac{1}{2}Nm\dot{x}^2(\boldsymbol{r},t) = (2\varepsilon_o\omega_p^2)^{-1}\partial_t \boldsymbol{P}(\boldsymbol{r},t) \cdot \partial_t \boldsymbol{P}(\boldsymbol{r},t), \quad (31\text{a})$$

$$\mathcal{E}_P(\boldsymbol{r},t) = \tfrac{1}{2}N\alpha x^2(\boldsymbol{r},t) = (2\varepsilon_o\omega_p^2)^{-1}\omega_r^2 \boldsymbol{P}(\boldsymbol{r},t) \cdot \boldsymbol{P}(\boldsymbol{r},t). \quad (31\text{b})$$

In these equations, $N$ is the number-density of the dipoles, $m$ is the mass of oscillating particles of charge $q$ associated with individual oscillators, $\alpha$ is the spring constant, and $x(\boldsymbol{r},t)$ is the displacement along the $x$-axis of each oscillating particle in response to the local $E_x$-field. The velocity of the particle along $x$ is denoted by $\dot{x}(\boldsymbol{r},t)$, the resonance frequency by $\omega_r = \sqrt{\alpha/m}$, the plasma frequency by $\omega_p = \sqrt{Nq^2/(\varepsilon_o m)}$, and polarization by $\boldsymbol{P}(\boldsymbol{r},t) = Nqx(\boldsymbol{r},t)\hat{\boldsymbol{x}}$. Recalling that, in the absence of loss mechanisms, the susceptibility of the host medium in the Lorentz oscillator model is

$$\varepsilon_o \chi(\omega) = \varepsilon_o \omega_p^2/(\omega_r^2 - \omega^2), \quad (32)$$

which yields a refractive index $n(\omega) = \sqrt{1 + \chi(\omega)}$, and considering that

$$\boldsymbol{P}(\boldsymbol{r},t) = (2\pi)^{-1}\int_{-\infty}^{\infty}\varepsilon_o \chi(\omega)\boldsymbol{E}(\omega)\exp\{i[\boldsymbol{k}(\omega)\cdot \boldsymbol{r} - \omega t]\}d\omega, \quad (33)$$

we may now evaluate the total EM energy of the system as follows:

$$\mathcal{E}(t) = \int_{-\infty}^{\infty}\{\tfrac{1}{2}\varepsilon_o \boldsymbol{E}(\boldsymbol{r},t)\cdot \boldsymbol{E}(\boldsymbol{r},t) + \tfrac{1}{2}\mu_o \boldsymbol{H}(\boldsymbol{r},t)\cdot \boldsymbol{H}(\boldsymbol{r},t)$$
$$+ (2\varepsilon_o\omega_p^2)^{-1}[\partial_t \boldsymbol{P}(\boldsymbol{r},t)\cdot \partial_t \boldsymbol{P}(\boldsymbol{r},t) + \omega_r^2 \boldsymbol{P}(\boldsymbol{r},t)\cdot \boldsymbol{P}(\boldsymbol{r},t)]\}dz$$
$$= \tfrac{\varepsilon_o}{8\pi^2}\iint_{-\infty}^{\infty}\{1 + n(\omega)n(\omega') + [(\omega_r^2 - \omega\omega')/\omega_p^2]\chi(\omega)\chi(\omega')\}\boldsymbol{E}(\omega)\cdot \boldsymbol{E}(\omega')$$
$$\times \exp[-i(\omega + \omega')t]\int_{-\infty}^{\infty}\exp\{i[k(\omega) + k(\omega')]z\}dz d\omega d\omega'$$
$$= \tfrac{\varepsilon_o}{4\pi}\iint_{-\infty}^{\infty}\{1 + n(\omega)n(\omega') + [(\omega_r^2 - \omega\omega')/\omega_p^2]\chi(\omega)\chi(\omega')\}\boldsymbol{E}(\omega)\cdot \boldsymbol{E}(\omega')$$
$$\times \exp[-i(\omega + \omega')t]\,\delta[k(\omega) + k(\omega')]\,d\omega d\omega'$$
$$= \tfrac{\varepsilon_o}{4\pi}\int_{-\infty}^{\infty}(dk/d\omega)^{-1}\{1 + n(\omega)n^*(\omega) + [(\omega_r^2 + \omega^2)/\omega_p^2]\chi(\omega)\chi^*(\omega)\}\boldsymbol{E}(\omega)\cdot \boldsymbol{E}^*(\omega)d\omega$$

$$= (4\pi Z_0)^{-1} \int_{-\infty}^{\infty} [\partial_\omega \omega n(\omega)]^{-1} \{1 + n^2(\omega) + [(\omega_r^2 + \omega^2)\omega_p^2/(\omega_r^2 - \omega^2)^2]\} \mathbf{E}(\omega) \cdot \mathbf{E}^*(\omega) \mathrm{d}\omega$$

$$= (4\pi Z_0)^{-1} \int_{-\infty}^{\infty} [\partial_\omega \omega n(\omega)]^{-1} \{2n^2(\omega) + [2\omega^2 \omega_p^2/(\omega_r^2 - \omega^2)^2]\} \mathbf{E}(\omega) \cdot \mathbf{E}^*(\omega) \mathrm{d}\omega$$

$$= (2\pi Z_0)^{-1} \int_{-\infty}^{\infty} [\partial_\omega \omega n(\omega)]^{-1} [n^2(\omega) + \tfrac{1}{2}\omega\, \partial_\omega \chi(\omega)] \mathbf{E}(\omega) \cdot \mathbf{E}^*(\omega) \mathrm{d}\omega$$

$$= (2\pi Z_0)^{-1} \int_{-\infty}^{\infty} [\partial_\omega \omega n(\omega)]^{-1} [n(\omega) + \omega\, \partial_\omega \sqrt{1 + \chi(\omega)}\,] n(\omega) \mathbf{E}(\omega) \cdot \mathbf{E}^*(\omega) \mathrm{d}\omega$$

$$= (2\pi Z_0)^{-1} \int_{-\infty}^{\infty} n(\omega) \mathbf{E}(\omega) \cdot \mathbf{E}^*(\omega) \mathrm{d}\omega. \tag{34}$$

The above expression is identical to that found in Eq.(27) for the total energy content of the light pulse inside a homogeneous and transparent dielectric slab of refractive index $n(\omega)$.

**8. Concluding remarks**. This paper has provided detailed explanations for some of the classical thought experiments pertaining to EM linear and angular momenta. Maxwell's equations, the Lorentz force law, special theory of relativity, and conservation laws of classical physics have been called upon to deduce the relations between the energy and the linear and angular momenta of EM wavepackets propagating in free space. In the context of the Balazs thought experiment, a detailed analysis of the EM energy and linear momentum inside a transparent host revealed the EM part of the momentum to be the same as that in free space divided by the group refractive index $n_g$ of the host medium. In the modern literature of EM momentum, the Balazs thought experiment is often cited as proof that, inside material media, the EM momentum of light is the Abraham momentum.[12]

**References**

1. J. A. Stratton, *Electromagnetic Theory*, McGraw-Hill, New York (1941).
2. L. Landau and E. Lifshitz, *Electrodynamics of Continuous Media*, Pergamon, New York (1960).
3. R. P. Feynman, R. B. Leighton, and M. Sands, *The Feynman Lectures on Physics*, Addison-Wesley, Reading (1964).
4. J. D. Jackson, *Classical Electrodynamics* (3rd edition), Wiley, new York (1998).
5. M. Mansuripur, *Field, Force, Energy and Momentum in Classical Electrodynamics* (revised edition), Bentham Science Publishers, Sharjah, UAE (2017).
6. L. Allen, S. M. Barnett, and M. J. Padgett, *Optical Angular Momentum*, Institute of Physics Publishing, Bristol, United Kingdom (2003).
7. M. Mansuripur, "Energy and linear and angular momenta in simple electromagnetic systems," Optical Trapping and Optical Micromanipulation XII, *Proceedings of SPIE* **9548**, 95480K~1-24 (2015).
8. M. Mansuripur, "Optical angular momentum in classical electrodynamics," *Physica Scripta* **92**, 065501 (2017).
9. S. Ramo, J. R. Whinnery and T. Van Duzer, *Fields and Waves in Communication Electronics* (2nd edition), Wiley, New York (1984).
10. A. W. Snyder and J. D. Love, *Optical Waveguide Theory*, Chapman and Hall, London (1983).
11. N. L. Balazs, "The energy-momentum tensor of the electromagnetic field inside matter," *Physical Review* **91**, 408-411 (1953).
12. R. Loudon, "Radiation Pressure and Momentum in Dielectrics," De Martini lecture, *Fortschritte der Physik* **52**, 1134-1140 (2004).
13. I. S. Gradshteyn and I. M. Ryzhik, *Table of Integrals, Series, and Products*, Academic Press, New York (1989).